\documentclass[%
 reprint,
 amsmath,amssymb,
 aps,
pra,
]{revtex4-2}

\usepackage{graphicx}
\usepackage{dcolumn}
\usepackage{bm}
\usepackage{hyperref}

\usepackage{xcolor}
\usepackage{wasysym}
\usepackage{physics}
\usepackage{siunitx}
\usepackage{float}

\begin{document}

\preprint{APS/123-QED}

\title{Nonlinear engineering of Hong-Ou-Mandel interference with structured light}

\author{Simon Darveau}
\email{simon.darveau@inrs.ca}
\author{Bienvenu Ndagano}%
\email{bienvenu.ndagano@inrs.ca}
\affiliation{Centre Énergie, Matériaux \& Télécommunications, Institut National de la Recherche Scientifique, 1650 Bd Lionel-Boulet, Varennes, QC J3X 1P7, Canada
}%


\date{\today}

\begin{abstract}
A pair of photons entering a Hong-Ou-Mandel interferometer will bunch or anti-bunch depending on the symmetry of their state in the measured degree of freedom, resulting in constructive or destructive interference. For photon pairs generated through spontaneous parametric down-conversion, this symmetry can be engineered through linear transformations and projections applied to the photon pairs, or by tailoring the pump field. Here, we adopt the latter approach to demonstrate that the pump's parity affects the shape and frequency of the transition from constructive to destructive interference. We demonstrate this concept theoretically and experimentally using a rotating structured pump mode encoded in the Hermite-Gaussian and Laguerre-Gaussian bases. In the former case, we show that the transition exhibits a rich structure unique to the mode family, resulting in enhanced sensitivity as the mode order increases. In the latter case, we demonstrate, as with NOON states, an integer enhancement in angular resolution by a factor proportional to the orbital angular momentum content, thereby opening the possibility of Hong-Ou-Mandel-enhanced rotation sensing.

\end{abstract}

\maketitle


\section{Introduction}
Two-photon interference at a beam-splitter, also known as the Hong-Ou-Mandel (HOM) effect, was first demonstrated in 1987 by Chung Ki Hong, Zhe Yu Ou, and Leonard Mandel \cite{hong_measurement_1987,bouchard_two-photon_2020}. When two independent, indistinguishable photons enter the two input ports of a 50:50 beam-splitter, they bunch at the two output ports, producing the characteristic "HOM dip" observed in coincidence detection between the outputs. Indistinguishability is central to a range of quantum systems and processes; HOM interferometry has been used to engineer quantum states \cite{zhang_engineering_2016,ulanov_loss-tolerant_2016,chen_polarization_2018,georgi_metasurface_2019}, characterize quantum systems \cite{He_2013,kim_two-photon_2016,senellart_2017,ollivier_hong-ou-mandel_2021, gao_high-speed_2022}, and process quantum information \cite{zhang_simultaneous_2017, stobinska_quantum_2019}. In photon-matter interactions, changes in indistinguishability have been used to extract information about a sample, enabling quantum imaging \cite{ndagano_quantum_2022,ibarra-borja_experimental_2020}, precision metrology \cite{polino_photonic_2020,singh_fast_2024, li_sample-half-inserted_2025}, and spectroscopy \cite{dorfman_hong-ou-mandel_2021,chen_entanglement-assisted_2022}.

Another aspect of HOM interferometry is its sensitivity to state symmetry \cite{bouchard_two-photon_2020,descamps_role_2026}; a symmetric two-photon input state bunches at the output port, while an antisymmetric state anti-bunches (producing a "HOM peak"). These two outcomes can also be viewed as constructive (anti-bunching) and destructive (bunching) two-photon interference. In the bosonic context of photons, constructive two-photon interference is associated with an antisymmetric state, and therefore with an entangled state. It has been shown that the transition from constructive to destructive interference can be associated with a linear transformation from one Bell state to another \cite{zhang_engineering_2016, gao_manipulating_2022}, and that this transition can be harnessed for birefringence imaging \cite{goncalves_quantum_2026} and entanglement distillation \cite{ndagano_entanglement_2019}.

The state symmetry can also be controlled in a nonlinear fashion; in spontaneous parametric downconversion, the symmetry of the generated photon pairs can be engineered by manipulating the pump's spatial parity \cite{yarnall_synthesis_2007}. Earlier work by Walborn \textit{et al.} presented an elegant realization using a structured pump mode \cite{walborn_multimode_2003}. Using a wire placed in a laser cavity, they tailored the pump beam to a first-order Hermite-Gaussian mode and showed that the (anti)bunching behaviour depends on the transverse orientation of the pump mode relative to the plane of propagation. In a different report, Walborn \textit{et al.} also showed that this control over the pump symmetry enables Bell state analysis using HOM interferometry \cite{walborn_optical_2003}.

In this work, we extend our control over pump symmetry and explore the properties of HOM interference with higher-order spatial modes. Using a spatial light modulator (SLM), we tailor the pump's spatial profile to generate higher-order Hermite- (HG) and Laguerre-Gaussian (LG) beams. For both modal families, the transition from spatial symmetry to antisymmetry can be achieved by rotating about the optical axis. We encode this rotation on the SLM and reveal rich structures in the HOM interferograms. On the one hand, we show that rotating a higher-order HG beam produces a HOM interferogram unique to the encoded pair of modal indices. We also show that increasing the modal order enhances angular sensitivity at the cost of angular range. On the other hand, the same rotation of a higher-order LG beam produces a sinusoidal HOM interference signal whose frequency increases with the number of quanta of orbital angular momentum. We note that this increase in frequency was shown by Barnett and Zambrini to increase angular resolution in a manner reminiscent of NOON states \cite{barnett_resolution_2006}.

\section{Theory}
We analytically investigate HOM interference using photon pairs generated through spontaneous parametric down-conversion (SPDC) pumped by a transversely rotating Hermite-Gaussian mode. We derive an expression for the coincidence rate at the output of a HOM interferometer as a function of the pump’s rotation angle. We then apply a transformation from the HG to the LG basis to express the HOM signal as a function of a rotating LG pump mode.
\begin{figure}[t] 
\centering
\includegraphics[width=0.65\linewidth]{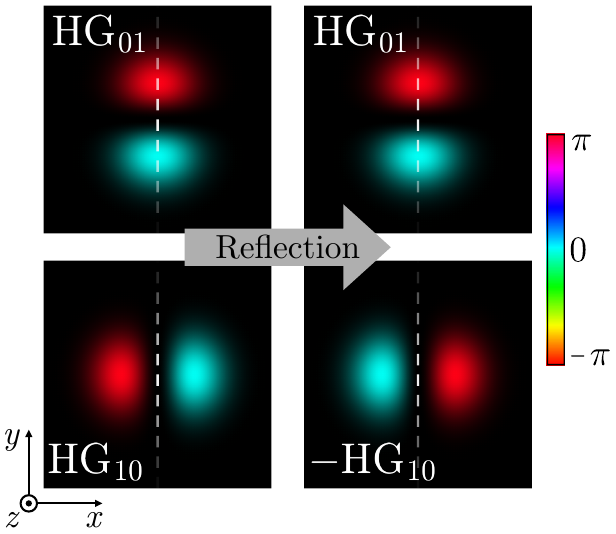}
\caption{Illustration of spatial symmetry. Normalized phase-amplitude maps of $\mathrm{HG}_{01}$ and $\mathrm{HG}_{10}$ modes are shown on the left. A reflection about the axis of symmetry (dashed line) leave $\mathrm{HG}_{01}$ unchanged, but adds a factor -1 to $\mathrm{HG}_{10}$ }
\label{fig:symmetry}
\end{figure}

It is useful to begin by defining the concept of symmetry in the present context. Consider the two modes, $\mathrm{HG}_{01}$ and $\mathrm{HG}_{10}$, shown in Fig.\ref{fig:symmetry}. We consider a reflection along the $x$-axis, about the line of symmetry indicated by the dashed white line. This operation leaves the $\mathrm{HG}_{01}$ mode unchanged, and we therefore classify it as symmetric. By contrast, the same reflection introduces a factor of $-1$ in the amplitude of the $\mathrm{HG}_{10}$ mode; we therefore classify it as antisymmetric. In \cite{walborn_multimode_2003}, it was shown shown that this reflection can be physically implemented by the reflecting surface of a beam-splitter. More generally, we can state:
\begin{equation} \label{parityHG}
\mathrm{HG}_{nm}(-x,y,z)=(-1)^n\,\mathrm{HG}_{nm}(x,y,z).
\end{equation}

\subsection{HOM interference of HG pump modes}
We consider a continuous-wave $\mathrm{HG}_{nm}$ pump beam incident on a nonlinear crystal, generating signal ($s$) and idler ($i$) photons via SPDC. Assuming a paraxial pump and monochromatic SPDC (enforced by narrow-bandwidth filters), it was shown in \cite{walborn_conservation_2005} that the entangled two-photon state generated is:
\begin{equation} \label{psi_nmCreator}
\ket{\psi_{nm}^{\mathrm{in}}}
=
\sum_{j,k,u,t=0}^{\infty}
C_{jkut}^{\,nm} \,
\hat{a}_{jk}^{\dagger}
\hat{b}_{ut}^{\dagger}
\ket{0},
\end{equation}
where $\hat{a}_{jk}^{\dagger}$ and $\hat{b}_{ut}^{\dagger}$ are creation operators that produce photons in paths $a$ and $b$ with spatial modes $\mathrm{HG}_{jk}$ and $\mathrm{HG}_{ut}$, respectively. To conserve spatial parity in both transverse directions, we also require.
\begin{align}
j+u \geq n, \qquad \mathrm{parity}(j+u)&=\mathrm{parity}(n), \label{eq:parity_x}\\
k+t \geq m, \qquad \mathrm{parity}(k+t)&=\mathrm{parity}(m). \label{eq:parity_y}
\end{align}
From SPDC, we have symmetry in the probability amplitudes of the product states, namely $C_{jkut}^{\,nm}=C_{utjk}^{\,nm}$. We can thus write:
\begin{equation}
\ket{\psi_{nm}^{\mathrm{in}}}
=
\frac{1}{2}
\sum_{j,k,u,t=0}^{\infty}
C_{jkut}^{\,nm}
\left(
\hat{a}_{jk}^{\dagger}\hat{b}_{ut}^{\dagger}
+
\hat{a}_{ut}^{\dagger}\hat{b}_{jk}^{\dagger}
\right)
\ket{0}.
\end{equation}

\begin{figure*}[t]
\centering
\includegraphics[width=\linewidth]{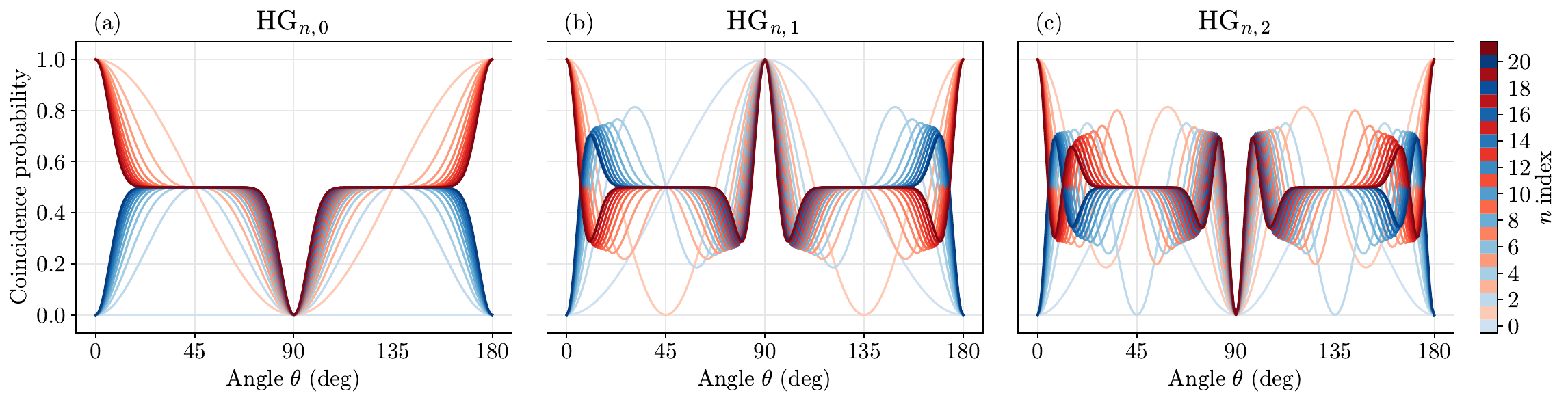}
\caption{(Color online) Angular dependence of HOM interference for several higher-order $\mathrm{HG}_{n,m}$ pump modes with $n$ ranging from 0 to 21 and $m=0$, $1$, and $2$ in (a), (b), and (c), respectively. A blue gradient is used for even $n$ indices, while a red gradient is used for odd $n$ indices.}
\label{fig:CoincProbTheo_m012}
\end{figure*}

The two photons then impinge on a balanced beam-splitter. Accounting for the effect of reflection on HG modes in Eq.~\eqref{parityHG}, the creation operators transform as follows: 
\begin{align}
\hat{a}_{jk}^{\dagger}
&\rightarrow
\frac{1}{\sqrt{2}}
\left[
(-1)^j \hat{c}_{jk}^{\dagger} \label{eq:adague}
+
\hat{d}_{jk}^{\dagger}
\right], \\ \label{eq:bdague}
\hat{b}_{ut}^{\dagger}
&\rightarrow
\frac{1}{\sqrt{2}}
\left[
\hat{c}_{ut}^{\dagger}
-
(-1)^u \hat{d}_{ut}^{\dagger}
\right],
\end{align}
where $c$ and $d$ denote the beam-splitter's output ports. Taking into account the parity conditions in Eqs.~(\ref{eq:parity_x}) and (\ref{eq:parity_y}), the state at the beam-splitter's output is:

\begin{widetext}
\begin{equation} \label{eq:MainResult}
\ket{\psi_{nm}^{\mathrm{out}}}
=
\frac{1}{4}
\sum_{j,k,u,t=0}^{\infty}
C_{jkut}^{\,nm}
\bigg[
(-1)^j\left(1+(-1)^n\right)
\left(
\hat{c}_{jk}^{\dagger}\hat{c}_{ut}^{\dagger}
+
\hat{d}_{ut}^{\dagger}\hat{d}_{jk}^{\dagger}
\right)
+
\left(1-(-1)^n\right)
\left(
\hat{c}_{jk}^{\dagger}\hat{d}_{ut}^{\dagger}
+
\hat{c}_{ut}^{\dagger}\hat{d}_{jk}^{\dagger}
\right)
\bigg]
\ket{0}.
\end{equation}
\end{widetext}

Equation~\eqref{eq:MainResult} shows that the photon statistics at the output ports of the HOM interferometer depend on the HG pump index $n$, which is associated with (anti-)symmetry in Eq.~\ref{parityHG}. This relation can be summarized as follows:
\begin{equation} \label{eq:conclusionBehavior}
\begin{cases}
\text{if } n \text{ is even} & \rightarrow \text{ photon bunching},\\[4pt]
\text{if } n \text{ is odd} & \rightarrow \text{ photon anti-bunching}.
\end{cases}
\end{equation}
Note that Eq.~\eqref{eq:MainResult} is independent of the index $m$ of the $\mathrm{HG}_{nm}$ pump beam, as expected, because the system is sensitive only to the spatial parity along the $x$, as discussed earlier.

\subsection{HOM interference of rotated HG pump modes}
Having established the effect of pumping the nonlinear crystal with any $\mathrm{HG}_{nm}$ mode, we now consider the pump's transverse rotation angle $\theta$. To this end, we express the rotated HG mode as a superposition of unrotated HG modes using the Wigner small-$d$ matrix formalism for transverse mode rotations, based on the $SU(2)$ representation developed in \cite{wolf_rotation_2008}. We map each HG mode onto a spin-like state, allowing us to describe transverse rotations using well-known Wigner small-$d$ matrix elements. For an HG mode of total order $N=n+m$, we introduce the following indices:
\begin{equation}
    j = \frac{N}{2},
    \qquad
    m_j = \frac{m-n}{2}.
\end{equation}
An HG mode rotated by an angle $\theta$ can then be expressed in terms unrotated HG modes of the same order $N$ as

\begin{equation} \label{eq:RotatedToNonRotatedHG}
    \ket{\mathrm{HG}_{n,m}(\theta)}
    =
    \sum_{m_j'=-j}^{j}
    d^{\,j}_{m_j',m_j}(2\theta)
    \ket*{\mathrm{HG}_{j-m_j', \,j+m_j'}},
\end{equation}
where $m_j=-j,-j+1,\ldots,j$ and $d^{\,j}_{m_j',m_j}$ are the Wigner small-$d$ matrix elements given by:
\begin{widetext}
\begin{equation}
d^{\,j}_{m_j',m_j}(2\theta)=
\sum_{k=k_{\mathrm{min}}}^{k_{\mathrm{max}}}
(-1)^{k-m_j'+m_j}
\frac{
\sqrt{
(j+m_j)!(j-m_j)!(j+m_j')!(j-m_j')!
}
}{
(j+m_j-k)!\,k!\,(j-k-m_j')!\,(k-m_j+m_j')!
}
(\cos\theta)^{2j+m_j-m_j'-2k}
(\sin\theta)^{2k-m_j+m_j'}. 
\end{equation}
\end{widetext}
The index $k$ ranges from $k_{\mathrm{min}}=\max(0,m_j-m_j')$ to $k_{\mathrm{max}}=\min(j+m_j,j-m_j')$. Using Eq.~\eqref{eq:MainResult}, the HOM coincidence probability for an arbitrarily rotated HG pump mode is obtained by summing the squared moduli of the Wigner small-$d$ matrix elements over all $\mathrm{HG}$ modes with odd $n$ (equivalently, odd $j-m_j'$). Thus, the coincidence probability, $P_{nm}^{\,\mathrm{coinc}}(\theta)$, associated with anti-bunching for an $\mathrm{HG}_{nm}$ pump mode rotated by an angle $\theta$ is given by
\begin{equation}\label{eq:probAnti}
P_{nm}^{\,\mathrm{coinc}}(\theta)
=
\sum_{\substack{m_j'=-j \\ j-m_j'\,\mathrm{odd}}}^{j}
\left|
d^{\,j}_{m_j',m_j}(2\theta)
\right|^2.
\end{equation}
Note that $P(\theta) \in \left[0, 1\right]$, with $P=0$ and $P=1$ correspond to complete bunching and anti-bunching, respectively. For $P = \tfrac{1}{2}$ we have equal probability of bunching and anti-bunching.

Figure~\ref{fig:CoincProbTheo_m012} shows the theoretical coincidence probabilities for $\mathrm{HG}_{nm}$ pump modes with $m=0,1,2$ as a function of the rotation angle $\theta$. As predicted by Eq.~\eqref{eq:MainResult}, the non-rotated $\mathrm{HG}_{n,m}$ modes ($\theta=0^\circ$ and $180^\circ$) exhibit photon bunching for even $n$ (blue shades) and photon anti-bunching for odd $n$ (red shades). At $\theta=90^\circ$, the HOM behavior is instead determined by the index $m$. This is because, for HG modes, a $90^\circ$ rotation swaps the indices. Thus, the index $n$ is replaced by $m$ in Eq.~\eqref{eq:MainResult}: (odd)even $m$ leads to photon (anti-)bunching. Interestingly, as the index $n$ increases, the transitions away from (anti-)bunching sharpen around $\theta = 0^\circ$ and $90^\circ$, and we observe the appearance of plateaus in between. This indicates an enhancement in sensitivity to angular rotation around these key points, at the cost of a reduced angular range.

\begin{figure}[t] 
\centering
\includegraphics[width=\linewidth]{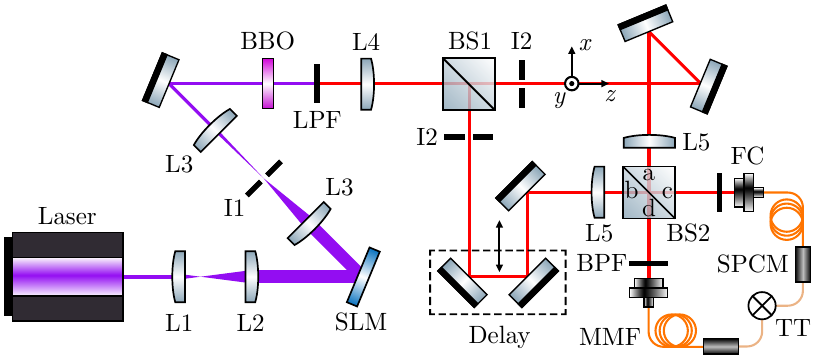}
\caption{(Color online) Schematic of the experimental setup to engineer HOM interference through spatial parity. SLM: spatial light modulator; BBO: $\beta$-barium borate crystal; LPF: longpass filter; BS: 50:50 beam-splitter; BPF: bandpass filter; MMF: multimode optical fiber \textcolor{black}{$(\diameter\!=\!25\,\mu\mathrm{m})$}; SPCM: single-photon counting modules; TT: time tagger; I: iris; $f_{1} \!=\! f_{4} \!=\!50\,\mathrm{mm}$; $f_{2}\!=\!150\,\mathrm{mm}$; $f_{3}\!=\!100\,\mathrm{mm}$; $f_{5}\!=\!500\,\mathrm{mm}$; $f_{\mathrm{FC}}\!=\!15\,\mathrm{mm}$; a, b: input ports of BS2; c, d: output ports of BS2}
\label{fig:setup}
\end{figure}

\subsection{HOM interference of LG pump modes}
Having established the Hong-Ou-Mandel interference behavior of rotating Hermite-Gaussian (HG) modes, we now investigate the corresponding behavior when the nonlinear crystal is pumped with Laguerre-Gaussian (LG) modes. To this end, we express a given $\mathrm{LG}_{p\ell}$ mode as a superposition of $\mathrm{HG}_{nm}$ modes with the same total order
$N = 2p + |\ell| = n+m$.
The corresponding HG and LG modal indices are related as follows:
\begin{equation}
n = p+\max(\ell,0),
\qquad
m = p+\max(-\ell,0).
\end{equation}
The decomposition can then be written as \cite{beijersbergen_astigmatic_1993}
\begin{equation} \label{eq:LGtoHG}
\ket{\mathrm{LG}_{n,m}}
=
\sum_{k=0}^{N}
i^k\, b(n,m,k)\,
\ket{\mathrm{HG}_{N-k,\,k}},
\end{equation}
where the expansion coefficients are given by
\begin{equation}
b(n,m,k)
=
\sqrt{
\frac{(N-k)!\,k!}
{2^N\,n!\,m!}
}
\frac{1}{k!}
\left.
\frac{\mathrm{d}^k}{\mathrm{d}t^k}
\left[
(1-t)^n(1+t)^m
\right]
\right|_{t=0}.
\end{equation}
Then, to determine the angular dependence, we rewrite Eq.~\eqref{eq:LGtoHG} for a rotated LG mode as
\begin{equation} \label{eq:RotatedLGtoRotatedHG}
\ket{\mathrm{LG}_{n,m}(\theta)}
=
\sum_{k=0}^{N}
i^k\,b(n,m,k)\,
\ket{\mathrm{HG}_{N-k,\,k}(\theta)},
\end{equation}
where each rotated HG mode is subsequently expanded in the non-rotated HG basis using Eq.~\eqref{eq:RotatedToNonRotatedHG}. 

By a rotational symmetry argument, one can correctly predict that pure $\mathrm{LG}_{p\ell}$ modes will exhibit equal probabilities of bunching and anti-bunching and will show no dependence on the rotation angle: $P_{p\ell}^{\,\mathrm{coinc}} = 1/2$. However, the same argument cannot be made for their superposition; we consider a rotated superposition of pure LG modes with conjugate orbital angular momentum (OAM) charge (often called petal mode): 
\begin{equation}\label{eq:petal}
\ket*{\mathrm{LG^+_{p,\ell}}}=\frac{1}{\sqrt{2}}\Big(\ket*{\mathrm{LG}_{p,\ell}(\theta)} +\ket*{\mathrm{LG}_{p,-\ell}(\theta)}\Big).
\end{equation}
Using Eqs.~(\ref{eq:probAnti}) and (\ref{eq:RotatedLGtoRotatedHG})  one finds that the coincidence probability (anti-bunching) when pumping with a rotated petal mode is given by:
\begin{equation} \label{eq:petalAngDependence}
P_{\mathrm{\ell}}^{\,\mathrm{coinc,+}}(\theta)
=
\begin{cases}
\sin^2(\ell\theta) & \text{if } \ell \text{ is even},\\[4pt]
\cos^2(\ell\theta)& \text{if } \ell \text{ is odd}.
\end{cases}
\end{equation}
Naturally, the radial index $p$ does not affect the angular coincidence probability, since the radial and azimuthal degrees of freedom are independent. However, we observe a sinusoidal oscillation whose frequency increases with OAM charge $\ell$. This would result in enhanced rotational sensitivity with increasing OAM.

\begin{figure}[b]
\centering
\includegraphics[width=\linewidth]{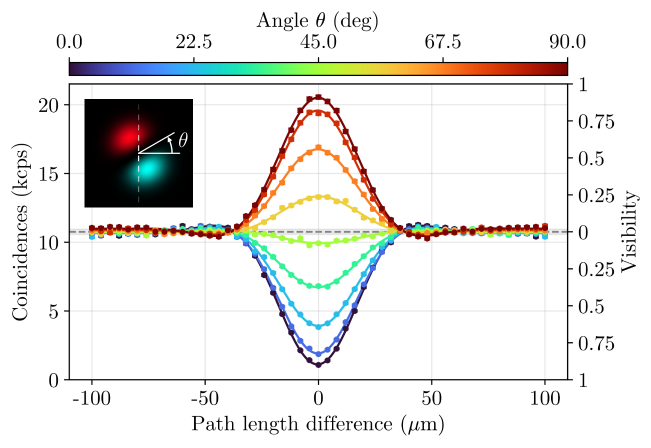}
\caption{(Color online) HOM signal as a function of relative time delay between the photon, and the rotation angle of the $\mathrm{HG}_{01}$ pump beam. The inset shows the theoretical intensity and phase profile of the rotating pump mode. The smooth lines are experimental fits. The dashed gray line represents the mean of the baseline measurements, and the shaded region corresponds to one standard deviation.}
\label{fig:Scans}
\end{figure}

\begin{figure*}[t]
\centering
\includegraphics[width=\linewidth]{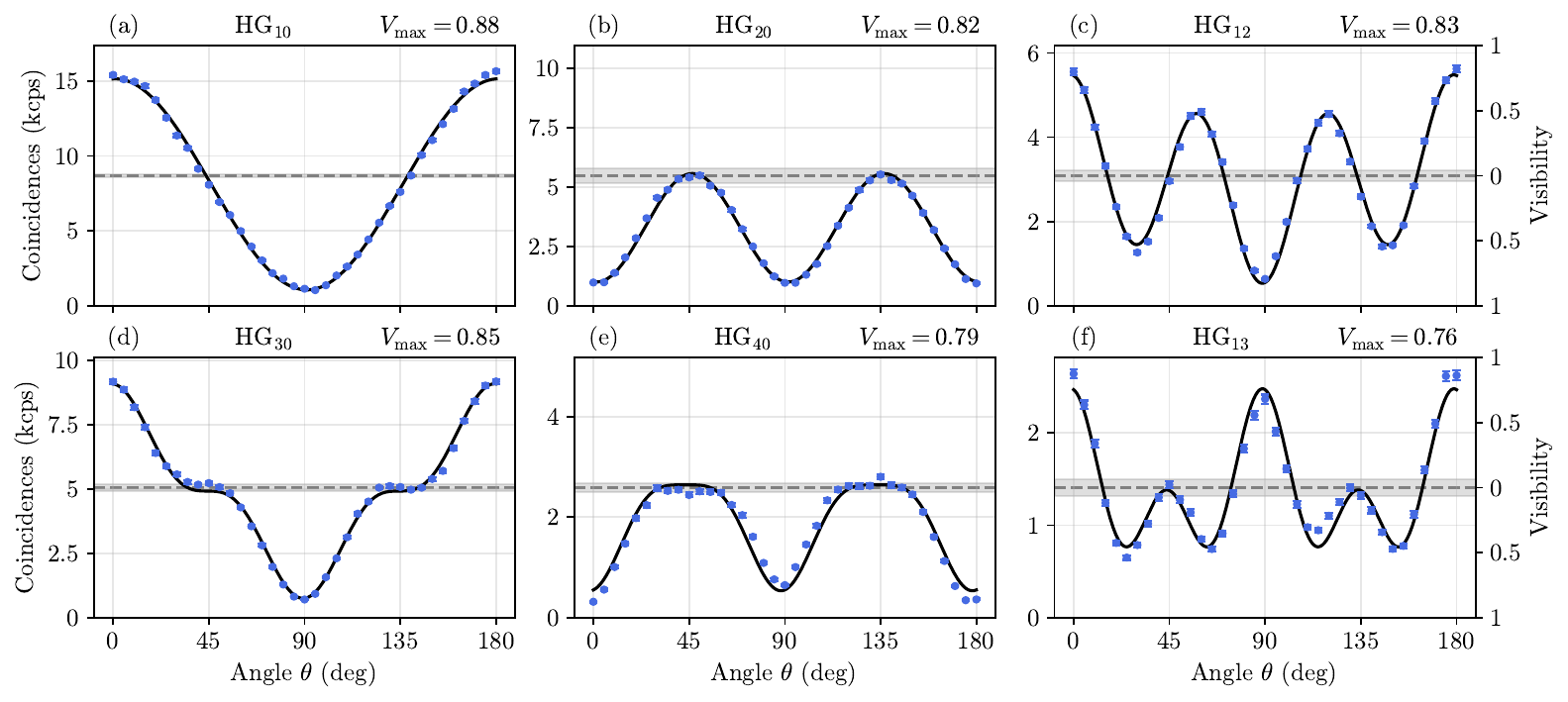}
\caption{(Color online) Angular dependence of HOM interference for several $\mathrm{HG}_{nm}$ pump modes. The pump is rotated in $\ang{5}$ increments over the range $[\ang{0},\ang{180}]$. At each angle, we acquire one baseline point (not shown), measured $100~\mu\mathrm{m}$ away from the dip or peak position, and one point at the dip or peak position (shown in color). The dashed gray line represents the mean of the baseline measurements, and the shaded region corresponds to one standard deviation. The solid lines are fits to the theoretical angular dependence given by Eq.~\eqref{eq:probAnti}. For (a)--(f), corresponding to $\mathrm{HG}_{10}$, $\mathrm{HG}_{20}$, $\mathrm{HG}_{12}$, $\mathrm{HG}_{30}$, $\mathrm{HG}_{40}$, and $\mathrm{HG}_{13}$ pump modes, respectively, the fitted functions are $\cos^2(\theta)$, $2\cos^2(\theta)\sin^2(\theta)$, 
$\left(12\cos^4(\theta)-18\cos^2(\theta)+7\right)\cos^2(\theta)$,
$\left(3\sin^4(\theta)+\cos^4(\theta)\right)\cos^2(\theta)$, $4\left(\sin^{4}{\left(\theta \right)} + \cos^{4}{\left(\theta \right)}\right) \sin^{2}{\left(\theta \right)} \cos^{2}{\left(\theta \right)}$ and $32\sin^8(\theta)-64\sin^6(\theta)+42\sin^4(\theta)-10\sin^2(\theta)+1$, respectively. The maximum fitted visibility is indicated in the upper-right corner of each plot.}
\label{fig:HGMosaic}
\end{figure*}
\section{Results \& Discussion}

The experimental setup is shown in Fig.~\ref{fig:setup}. A 200 mW continuous-wave laser operating at 405 nm was magnified and shaped using a phase-only hologram displayed on a  Pluto UV099B Holoeye SLM. The tailored beam was filtered in the first-order diffraction with an approximate power of 30 mW, and incident on a 1-mm-thick $\beta$-barium borate (BBO) crystal to produce degenerate photon pairs at 810 nm via Type-I phase matching. Signal and idler photons were deterministically separated using a beam-splitter and a pair or irises, post-selecting conjugate transverse momenta. The path of the two photons were recombined on a beam-splitter (BS2), with a motorized delay stage used to control their relative arrival time. Note that the total number of reflections experience experienced by both photons before BS2 have the same parity. Band-pass filters centered at 810 nm with a 10-nm bandwidth were placed in front of the fiber couplers connected to multi-mode fibers.

Figure~\ref{fig:Scans} shows the characterization of HOM interference for a rotated $\mathrm{HG}_{01}$ pump beam, thereby verifying the bunching and anti-bunching behavior predicted above. The error bars in all our results are estimated, using Poisson statistics, as the square root of the coincidence counts. We evaluated the  HOM interference visibility following Ref.~\cite{weihs_two-photon_1996}:
\begin{equation}
    V = \abs{\frac{C_{\mathrm{out}}-C_{\mathrm{in}}}{C_{\mathrm{out}}}},
\end{equation}
where $C_{\mathrm{in}}$ denotes the coincidence count rate at the dip or peak position, while $C_{\mathrm{out}}$ denotes the coincidence count rate outside the interference region -- we measured it at a $100~\mu\mathrm{m}$ path-length difference. We measured dip and peak visibilities of 90\% and 87\%, respectively.

\begin{figure*}[t]
\centering
\includegraphics[width=\linewidth]{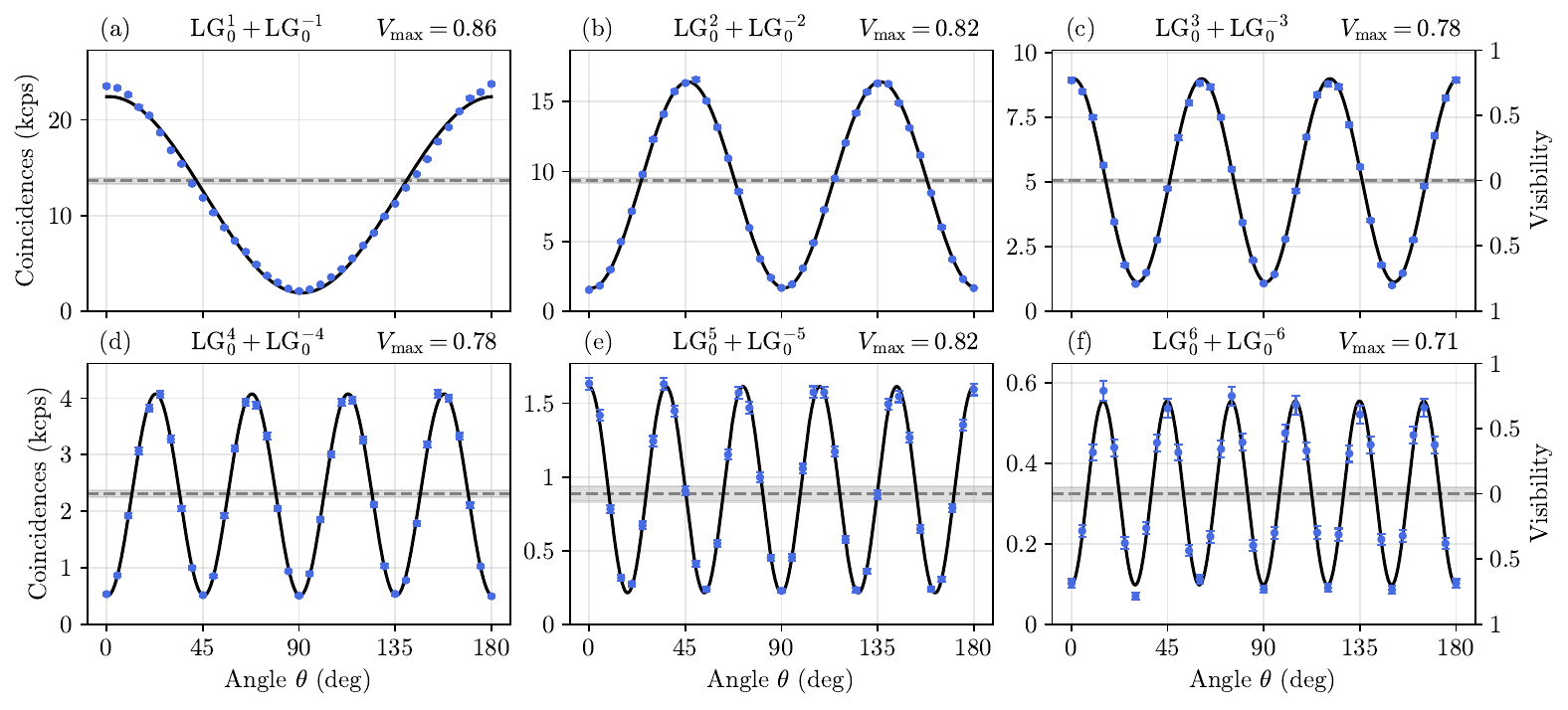}
\caption{(Color online) Angular dependence of HOM interference for several $\mathrm{LG}_{0}^{\ell}+\mathrm{LG}_{0}^{-\ell}$ petal beam pump modes. The pump is rotated in $\ang{5}$ increments over the range $[\ang{0},\ang{180}]$. At each angle, we acquire one baseline point (not shown), measured $100~\mu\mathrm{m}$ away from the dip or peak position, and one point at the dip or peak position (shown in color). The dashed gray line represents the mean of the baseline measurements, and the shaded region corresponds to one standard deviation. The solid lines are fits to the theoretical angular dependence given by Eq.~\eqref{eq:petalAngDependence}. For (a)--(f), corresponding to petal-beam pump modes with $\ell=1,2,3,4,5,6$, respectively, the fitted functions are $\cos^2(\ell\theta)$ for odd $\ell$ and $\sin^2(\ell\theta)$ for even $\ell$. The maximum fitted visibility is indicated in the upper-right corner of each plot.}
\label{fig:LGMosaic}
\end{figure*}

Next, we verified the transition from bunching to anti-bunching as a higher-order $\mathrm{HG}_{nm}$ pump mode is rotated, demonstrating the angular dependence of the HOM interference described by Eq.~\eqref{eq:probAnti}. The results shown in Fig.~\ref{fig:HGMosaic} show, up to $N=4$, excellent agreement with theory, with each mode exhibiting a unique interference pattern. It is worth noting that because $\mathrm{HG}_{mn}$ and $\mathrm{HG}_{nm}$ are related by a $90^\circ$ rotation, we do not evaluate modes with swapped indices. We also maintain high HOM (anti-)bunching visibility ($>75\%$) at $\ang{0}$ and $\ang{90}$, for all higher-order modes evaluated, well above the 50\% classical limit. We also observe the predicted narrowing of the interference signal around $\theta = \ang{0} \textrm{ and } \ang{90}$, showing an enhancement in sensitivity around that region. However, due to the finite numerical aperture, we lose coincidence counts with higher mode order, thereby limiting the range of modes available. 

Lastly, Fig.~\ref{fig:LGMosaic} shows similar measurements for higher-order petal modes up to $N=6$, where we observe an enhancement in oscillation frequency by a factor of $\ell$, in excellent agreement with theory. Here too, we maintain high visibility ($>70\%$) beyond the classical limit, but the range of available modes is limited as coincidence counts decrease with increasing mode order, as with HG modes. Nevertheless, we demonstrate enhanced sensitivity in rotation sensing, analogous to that observed for NOON states \cite{barnett_resolution_2006}, and consistent with a previous observation that projected SPDC photons onto higher-order modes using spatial light modulators and single-mode fibers \cite{hiekkamaki_high-dimensional_2021}.

An unexpected outcome of this investigation has been the use of multi-mode fibers to couple the photons at the output of the HOM interferometer. Generally, one would prefer single-mode fibers to project onto the same spatial mode, thereby ensuring spatial indistinguishability and maximizing visibility. Here, this approach would not work because SPDC parity conservation rules in \ref{eq:parity_x} and \ref{eq:parity_y} does not allow the generation of a pair of Gaussian modes accepted by single-mode fibers, with a higher-order pump beam; we would have $j=k=u=t=0$, with $j+u<n$ and $k+t<m$ for $m,n \geq 1$. The observed interference is therefore a result of multi-mode HOM interference.

\section{Conclusion}
We investigated the structure of Hong-Ou-Mandel interference for photon pairs generated via spontaneous parametric down-conversion (SPDC) pumped with higher-order Hermite-Gaussian and Laguerre-Gaussian modes. We digitally controlled the pump structure using a spatial light modulator (SLM), enabling the implementation of controlled rotations. When pumping with a Hermite-Gaussian mode, we observed that each mode family produces a distinctive signature as a function of the pump's rotation angle, corresponding to a transition from spatial symmetry to anti-symmetry. For higher-order Laguerre-Gaussian modes, we likewise measured distinctive signatures in superposition states known as petal beams. As with the Hermite-Gaussian modes, each mode order exhibits a distinctive signature under rotation. Notably, both sets of modes demonstrate enhanced sensitivity to rotation, suggesting the potential for quantum-enhanced Hong-Ou-Mandel sensing. Furthermore, the transition from a Hong-Ou-Mandel dip to a peak can be associated with a transition between symmetric and antisymmetric Bell states. Kaushik \textit{et al.} have recently shown polarization-encoded Bell state engineering through geometric phase control of the pump, and its applicability to HOM-enabled quantum sensing \cite{kaushik_quantum_2026}. The digital control we have demonstrated over HOM interference may also pave the way for using HOM interferometry as a physical computational layer.

\begin{acknowledgments}
The authors acknowledge financial support of the Quebec's MEIE and the NSERC Alliance Consortia Quantum Grant (QuEnSI). SD acknowledges financial support from NSERC's Alexander Graham Bell Canada Graduate Scholarships program. 
\end{acknowledgments}

\bibliographystyle{apsrev4-2}
\bibliography{apssamp}

\end{document}